\documentclass[11pt,epsf,amssymb,qsymbols]{article}
\usepackage[dvips]{graphicx,rotating}
 \usepackage{epsfig, amssymb}
\usepackage{tabularx}
\usepackage{array}
\usepackage{graphics}
\usepackage{graphicx}
\usepackage{psfrag}
\usepackage{epsfig}
\usepackage{amsmath}
\usepackage{amssymb}
\usepackage{ulem}
\usepackage{setspace}
\usepackage{rotating}
\usepackage{colortbl}
\usepackage{tabularx}
\usepackage{longtable}
\usepackage{multirow}
\usepackage{hyperref}

\newcommand{\nn}{{\nonumber}}

\newcommand{\mumu}{\ifmmode {\mu^+\mu^-} \else ${\mu^+\mu^-} $ \fi}

\newcommand{\ba}{\begin{array}}
\newcommand{\ea}{\end{array}}
\newcommand{\bc}{\begin{center}}
\newcommand{\ec}{\end{center}}
\newcommand{\beq}{\begin{eqnarray}}
\newcommand{\eeq}{\end{eqnarray}}
\newcommand{\bes}{\begin{eqnarray*}}
\newcommand{\ees}{\end{eqnarray*}}
\newcommand{\Kz}{\ifmmode {\rm K^0_s} \else ${\rm K^0_s} $ \fi}
\newcommand{\Zz}{\ifmmode {\rm Z} \else ${\rm Z } $ \fi}
\newcommand{\qqbar}{\ifmmode {\rm q\bar{q}} \else ${\rm q\bar{q}} $ \fi}
\newcommand{\ccbar}{\ifmmode {\rm c\bar{c}} \else ${\rm c\bar{c}} $ \fi}
\newcommand{\bbbar}{\ifmmode {\rm b\bar{b}} \else ${\rm b\bar{b}} $ \fi}
\newcommand{\xxbar}{\ifmmode {\rm x\bar{x}} \else ${\rm x\bar{x}} $ \fi}
\newcommand{\rphi}{\ifmmode {\rm R\phi} \else ${\rm R\phi} $ \fi}

\begin{document}

\begin{flushright}
\begin{tabular}{l}
{\small\tt LAL 12-212}\\
{\small\tt LPT 12-58}
\end{tabular}
\end{flushright}
\begin{center}
\vskip 2.30cm\par
{\par\centering \textbf{\LARGE  
\bf Proposal to study $B_s \to \overline{D}_{sJ}$ transitions  }}\\
\vskip 1.25cm\par
{\scalebox{.84}{\par\centering \large  
\sc D.~Be\v{c}irevi\'c$^a$, A.~Le Yaouanc$^a$, L.~Oliver$^a$, J-C.~Raynal$^a$,  P.~Roudeau$^b$, J.~Serrano$^c$ }}
{\par\centering \vskip 0.3 cm\par}
{\sl \small
$^a$~ Laboratoire de Physique Th\'eorique, CNRS/Univ. Paris-Sud 11 (UMR 8627)\\
91405 Orsay, France}\\
{\par\centering \vskip 0.3 cm\par}
{\sl \small
$^b$ Laboratoire de l'Acc\'el\'erateur Lin\'eaire, Univ. Paris-Sud 11, CNRS/IN2P3 (UMR 8607)\\
91405 Orsay, France}\\
{\par\centering \vskip 0.3 cm\par}
{\sl \small
$^c$ Centre de Physique des Particules de Marseille, Univ. Aix-Marseille, CNRS/IN2P3\\
13288 Marseille, France}\\

{\vskip 1.05cm\par}
\end{center}

\vskip 0.55cm
\begin{abstract}
We propose that some puzzles in semileptonic decays of $B$ mesons to the broad $\overline{D}^{**}$ states could be clarified by studying at LHCb the corresponding decays with strange mesons $B_s^0 \to D_{s0}^-$. In particular, we point out that the non-leptonic decay $B_s^0 \to D_{s0}^- \pi^+$ and the like, being Class I decays (where factorization is expected to hold), could be a first step in this direction. The interpretation of results in both semileptonic and non-leptonic decays will presumably be easier due to the narrowness of the $D_{s0}^-$ state. On the other hand, we make a careful and detailed study of the experimental and theoretical situation in the case of the wide non-strange $\overline{D}^{**}$ case, and we update previous analyses.
\end{abstract}
\vskip 2.cm

\section{Motivation}

The long-standing problem of weak transitions between $B$ and the broad $L=1$  ($j=1/2$) states $\overline{D}^{\ast\ast}$ remains interesting to elucidate for at least two reasons:~\footnote{Note that we use the spectroscopic labels related to the heavy quark limit in which the angular momentum of the light degrees of freedom $j$ is a good quantum number. In that limit $L=1$ corresponds to both $j^P=(1/2)^+$ and $j^P=(3/2)^+$ doublets of heavy-light mesons. The $(1/2)^+$-doublet is denoted as $[D_{0}^\ast,D_{1}^\prime ]$, while the $ (3/2)^+$-doublet is referred to as $[D_{1}, D_{2}^\ast ]$. The states with the strange valence  quark are distinguished by an extra index ``$s$". } 

\begin{enumerate}
\item A lot of {\sl theoretical} effort has been devoted to understand these transitions by using several different approaches.  
 
\item A considerable {\sl experimental} effort to measure the corresponding quantities lead to controversies: the experiments seemed to disagree among themselves and/or with theory.
\end{enumerate}
In the following,  we explain the current situation in Part~\ref{sec:difficulties}, by discussing the theoretical expectations, experimental results and by comparing theory with experiment. We distinguish between the semileptonic and non-leptonic decays.  In Part \ref{sec:proposal} we propose a way to clarify the puzzles by studying the strange $D_s^{\ast\ast}$-states which happen to be narrow.

\part{\underline{\Large Difficulties with $B\to \overline D^{\ast\ast}$ weak decays}\label{sec:difficulties}}

Many papers and notes have been devoted to the above problem. The issues have been discussed and summarized some years ago in refs.~\cite{memorino1,memorino2} to which we refer for complementary information and references.

\section{The broad $L=1$ ($j=1/2$) $c (\bar u,\bar d)$ $D^{\ast\ast}$ states} 
\label{sec:broad}
There are two states with $L=1$ ($j=1/2$): one with $J^P=0^{+}$ ($D_0^\ast$), the other with $J^P=1^{+}$ ($D_1^\prime$). Both are expected to be broad, because of the strong $S$-wave decays to 
$D^{(\ast)} \pi$, and the fact that their mass is expected to be notably above the $D^{(\ast)} \pi$  thresholds.

They have been most clearly observed in the non-leptonic $B \to  \overline{D}^{\ast\ast} \pi$ decays wherefrom their properties, like widths and masses, have been established. Although the semileptonic decay rates are much larger than the non-leptonic ones, the number of observed events is in a reversed proportion as we explain below. 

The decay rates of the two $D^{\ast\ast}$ states into $D^{(\ast)} \pi$ are identified as their total widths, which is roughly expected from simple quark model calculations~\cite{jugeau}. The identification of the very broad bumps in 
$D^{(\ast)} \pi$ 
with the expected $D^{\ast\ast}$ states  is plausible although (i) the identification of  very broad resonances is not safe, and (ii) the observed discrepancies between the predicted and observed $D^{\ast\ast}_q$ states made the  $c \bar q$-interpretation questionable ($q$ being either $u,d$ or $s$ quark). We will briefly return to the latter point in Sec.~\ref{sec:comparison}.

Similar to the above-mentioned bumps were observed in the semileptonic $B\to \overline{D}^{(\ast)}\pi \ell^+\nu_{\ell}$ decays, but not with an accuracy allowing to determine the resonance's features independently. Rather, one uses the $D^{\ast\ast}$ properties found in $B \to  \overline{D}^{\ast\ast} \pi$ as input in order to estimate the semileptonic decay rate. 

Theoretically, however, the semileptonic decays are simpler to describe and require less assumptions than the non-leptonic ones, and we will discuss them in that order.

\section{Theoretical predictions for $B \to \overline{D}^{\ast\ast}$ IW functions}
\label{sec:theoretical}
In the heavy quark limit  for the $c$ and $b$ quarks all the form factors governing 
$B \to \overline{D}^{\ast\ast}\ell^+\nu_{\ell}$ decays are related by simple relations and  proportional to one of the two Isgur-Wise (IW) functions,  $\tau_{1/2}(w)$ for the final hadron belonging to the $j=1/2$ doublet, or $\tau_{3/2}(w)$ for $D^{\ast\ast}$ being one of the mesons from the $j=3/2$ doublet. These functions parameterize the non-perturbative QCD dynamics of the vector or axial current matrix elements~\cite{Isgur:1990jf} as, for example,
\begin{eqnarray}
\langle 0^+~|A_{\mu}|~0^- \rangle&=&-
\frac {1}{\sqrt{v_0v_0'}}(v_{\mu}-v_\mu^\prime)~\tau_{1/2}(w)\,,\nn \\
\langle 2^+~|A_{\mu}|~0^- \rangle&=&\frac {\sqrt{3}}{2}
\frac {1}{\sqrt{v_0v_0'}}\left[ (1+w) \epsilon_{\mu \nu}^*v^{\nu}-v_{\mu}^\prime v^{\nu}v^{\rho}\epsilon_{\nu\rho}^\ast \right]~\tau_{3/2}(w)\,,
\end{eqnarray}
where $v,v'$ are the velocity vectors of the initial and final mesons, $\epsilon_{\mu\nu}$ is the polarisation tensor of the $2^+$-state, and $w=v\cdot v^\prime$. The normalisation of states is $(2 \pi)^3~\delta(\vec v -\vec v^\prime)$.

The argument of $\tau_{j}(w)$ varies between $1\leq w \lessapprox  1.3$, as it can be easily seen from 
\begin{equation}
w=\frac {m_B^2+m_{D^{\ast\ast}}^2-q^2} {2~m_B~m_{D^{\ast\ast}}}\,,
\end{equation}
for $q_{\rm min}^2=m_\ell^2 \approx 0$, and $q^2_{\rm max}= (m_B-m_{D^{\ast\ast}})^2$. For the non-leptonic decays $q^2=m_{\pi}^2$ is fixed and corresponds to $w\approx 1.3$. Importantly, $\tau_{1/2}(w)$ is known to be a slowly varying function of $w$, and it is a common practice to focus on its normalization at zero recoil $w=1$, namely $\tau_{1/2}(1)$.  For example, in ref.~\cite{morenas} it was found that $\tau_{1/2}(w) = \tau_{1/2}(1)[ 1- 0.83 (w-1) + \dots]$.
\vskip 0.5cm
\subsection{Inclusive sum rules}
\label{sec:SR}
A useful constraint concerning the values of $\tau_j(1)$ is provided by what we can call Bjorken-like or inclusive sum rules, which are not to be confused with the ``QCD Sum Rules" \`a la SVZ~\cite{SVZ} in that they do not pretend to go beyond equalling the sum over all states of suitable quantum numbers to the result obtained by employing the operator product expansion (OPE). They in fact reflect the duality with free quarks.
One of the most famous such sum rule is the so called Uraltsev SR~\cite{uraltsev}, 
\begin{eqnarray}
\sum_{n }  |\tau_{3/2}^{(n)}(1)|^2-|\tau_{1/2}^{(n)}(1)|^2=\frac{1}{4}\,,
\end{eqnarray}
with $n$ labeling possible radial excitations ($n=0$ being the ground state).
Focusing only onto the ground states suggests the inequality  $|\tau_{1/2}(1)| < |\tau_{3/2}(1)|$, which is also confirmed by the similar sum rule studied in ref.~\cite{interesting}.  This is obviously not a theorem, but relies on assumption that the lowest state dominates in each channel. The right hand side, $1/4$ may seem a small difference, but since $\vert \tau_{1/2}(1)\vert^2$ is a small number, the ratio $|\tau_{3/2}(1)/\tau_{1/2}(1)|$ is rather large,
\begin{eqnarray}
{    \left| \tau_{3/2}(1)\right|^2 \over \left| \tau_{1/2}(1)\right|^2 }=
1+\frac{1/4}{|\tau_{1/2}(1)|^2}\,,
\end{eqnarray}
when considering the lowest states only.
This tendency is observed in actual theoretical calculations, except in the QCD sum rule calculation of ref.~\cite{paver1}. In the semileptonic rates it is further exacerbated by the kinematic factors.

\vskip 0.5cm
\subsection{Lattice QCD predictions}
\label{sec:lattice}
The only method allowing to compute these form factors, strictly based on QCD, is the method of numerical simulations of QCD on the lattice. 
The first calculation of $\tau_{1/2}(1)$ has been made in ref.~\cite{blossier05} and then extended and improved in ref.~\cite{wagner09}, where the computation is made by including the $N_{\rm f}=2$ flavors of dynamical (``{\it sea}") quarks. The results of ref.~\cite{wagner09}, obtained at a single lattice spacing, exhibit a negligible dependence on the light quark mass and read:
\begin{eqnarray}\label{eq:latres}
\tau_{3/2}(1)= 0.528(23)\,,\qquad  \tau_{1/2}(1) = 0.297(26)\,,
\end{eqnarray} 
where the errors do not include the discretization nor the finite volume effects. Note also that one cannot easily calculate these form factors  away from $w=1$ on the lattice.
\vskip 0.5cm
\subsection{Quark model predictions}
\label{sec:quark}
Familiar opinions that ``{\sl any model would do}" or that ``{\sl you may get anything you want by choosing a suitable model}" come from disregarding the necessary careful discussions which allow to estimate the overall merits of respective models by consideration of the largest possible set of phenomenological data and of theoretical consistency and inputs.
 
There is no perfect model, other than QCD, but there are definitely bad models and more satisfactory ones. One necessary general feature is that for heavy-light systems they should be relativistic. As to external motion of hadrons one can use the Bakamjian-Thomas (BT) approach which provides a definite way to define states in motion starting from states at rest by constructing an explicit Poincar\'e algebra.
A particular case is obtained by performing boosts to the infinite momentum frame, which gives the familiar null-plane formalism. Covariance of current matrix elements is ensured in the heavy mass limit only. Note that the above inclusive sum rules, required by QCD, are exactly satisfied by the BT quark model approach. 

Within the BT quark model approach the difference between $\tau_{3/2}(1)$ and $\tau_{1/2}(1)$ 
comes from the Wigner rotations of the light spectator quark, which acts differently for $j=1/2$ and for $j=3/2$ states. One finds that the difference $|\tau_{3/2}(1) |-  |\tau_{1/2}(1) |$ is positive and large~\cite{morenas}. 

In addition to the quark model framework, one also has to choose a (necessarily relativistic) potential model  to fix  
the wave functions at rest.~\footnote{In a very extensive work, H. Cheng et al. \cite{cheng} have made predictions for the transitions to the $D^{\ast\ast}$-states in the null-plane formalism, including the finite $m_{b,c}$ effects, which is quite useful. However, to be conclusive, a necessary step in this approach, which remains to be done, would be to systematically deduce  the wave functions from a relativistic potential model constrained by the spectrum.} The guiding principle in choosing the potential is obviously the requirement to describe as broad range of observed hadrons as possible. 
In that respect, the standard Godfrey-Isgur (GI) potential model provides the best description of the whole spectroscopy. By using the wave functions fixed by the GI potential model, the BT approach leads to the following results:
\begin{eqnarray}\label{eq:BTGI}
\tau_{3/2}(1)\simeq  0.54\,,\qquad  \tau_{1/2}(1) \simeq 0.22\,.
\end{eqnarray}
The agreement with the results of lattice calculations~(\ref{eq:latres}), which have been produced much later, is striking. The suppression of $\tau_{1/2}(1)$ with respect to $\tau_{3/2}(1)$ could be even stronger if other potentials (other than GI) are chosen, while $\tau_{3/2}(1)$ remains stable. We do not quote errors to the above results because there is no clearly admitted definition of errors in the quark models, unlike in the well defined method of lattice QCD.
For instance, it would not make much sense to make an arbitrary variation of parameters without taking 
into account the whole set of possible phenomenological applications, most of which depend on additional modeling.

Before continuing we would like to emphasize the consistency of the results obtained in the static limit of QCD on the lattice with the results obtained by using the BT framework with a suitable potential model. Such an agreement is not just a matter of luck. A similar agreement has been observed in a very detailed manner for the distribution of the axial, scalar and vector charges in the static-light mesons with either $L=0$, or $L=1$~\cite{Becirevic:2011cj}. The advantage of quark models is that one can easily calculate the $w$-dependence of $\tau_{1/2,3/2}(w)$, needed when computing the branching ratios, and get moreover an intuitive insight.  

\vskip 0.5cm
\subsection{QCD sum rules approach to form factors}
\label{sec:QCDSR}
The results from QCD sum rules are less safe and less intuitive, and the results for $\tau_j(1)$ presented so far in the literature do not agree among themselves. A major concern is that the results depend quite strongly on the choice of the interpolating field for the $D^{\ast\ast}$-states.

Results of the first calculations presented in refs.~\cite{paver1,paver2},~\footnote{Result for $\tau_{3/2}(1)$ is read from the plot in ref.~\cite{paver1}, while the result for $\tau_{1/2}(1)$ was presented in ref.~\cite{paver2}.}
\begin{eqnarray}\label{eq:paver}
\tau_{3/2}(1)\sim   0.25\,,\qquad  \tau_{1/2}(1) \simeq 0.35(8)\,,
\end{eqnarray}
clearly challenge the hierarchy $|\tau_{1/2}(1)| < |\tau_{3/2}(1)|$. A little later another QCD sum rules computation resulted in~\cite{dai},  
\begin{eqnarray}\label{eq:dai}
\tau_{3/2}(1)\simeq  0.43(8)\,,\qquad  \tau_{1/2}(1) \simeq 0.13(4)\,,
\end{eqnarray}
arguing that the usual local scalar interpolating field operator does not lead to a satisfactory sum rule, due to a lack of perturbative contribution. To circumvent the problem they used the operators with covariant derivative instead. 
It must be stressed that the quoted ``errors" in eqs.~(\ref{eq:paver},\ref{eq:dai}) 
are not errors in the usual sense of indicating a possible deviation from the true value. They merely indicate the variation of the result within the chosen range for the continuum threshold. Therefore, one should neither consider $\tau_{1/2}(1) \simeq 0.13(4)$ as being incompatible with the result in eq.~(\ref{eq:paver}), nor incompatible with the values given in eqs.~(\ref{eq:latres},\ref{eq:BTGI}). The difference between the values in eqs.~(\ref{eq:paver}) and (\ref{eq:dai}) could be viewed as an indicator of a possible uncertainty of the method.  What is to be actually retained from the results of ref.~\cite{dai} is that the hierarchy is similar to the one found in the lattice QCD and in the quark model discussed above.~\footnote{Results we quote in eq.~(\ref{eq:dai}) are obtained after converting the values from ref.~\cite{dai} to our definitions of Isgur-Wise functions, namely, $\tau_{3/2}(1)= \tau(1)/\sqrt{3}$ and $\tau_{1/2}(1)=\zeta(1)/2$, where $\tau(1)$ and $\zeta(1)$ are defined in ref.~\cite{leib}.
}

\subsection{Phenomenology with $\tau_{1/2}(1)$ and $\tau_{3/2}(1)$} \label{phenomenological}
From the above discussion we see that there is a growing evidence that the Uraltsev sum rule is well respected by the actual values for the IW functions involving the $n=0$ $D^{\ast\ast}$-states at $w=1$, and that $\tau_{1/2}(1)< \tau_{3/2}(1)$. Of course the discussion so far has been restrained to the heavy quark limit of QCD. The impact of the corrections arising from the finiteness of the heavy quark mass has not been much discussed in the literature and there is no available lattice QCD result that would help us assess the size of these corrections. An early careful estimate of these corrections within a systematic HQET expansion of ref.~\cite{leib} suggests that they are small. Therefore, in what follows we will use the results for the form factors obtained in the static limit of QCD to compute the decay widths, but in the computation of the phase space we will use the physical meson masses. 

\subsubsection{Semileptonic decays in theory} The branching ratio of the semileptonic $B$-decay to a $j^P=(1/2)^+$ state should be very small compared to the decay to a $j^P=(3/2)^+$ meson. A suppression due to the IW functions
\begin{eqnarray}
\frac {|\tau_{1/2}(1)|^2} {|\tau_{3/2}(1)|^2} \simeq 0.17\,, 
\end{eqnarray}
is further enhanced by the phase space suppression (c.f. ref.~\cite{morenas}), and the suppression becomes one order of magnitude. Note that the decay to $D_1$ is less reliable because at $w=1$ its amplitude is zero. 

Using the results of the quark model calculation in the BT formalism with the GI potential model one has~\cite{morenas}:
\begin{eqnarray}
{\cal B}(B^0_d\to {D}_2^{\ast -}\ell \nu) &\simeq& 0.7 \times 10^{-2} \,,\nn\\
{\cal B}(B^0_d\to{D}^{-}_{1\ (3/2)}\ell \nu)&\simeq& 0.45 \times  10^{-2} \,,\nn \\
{\cal B}(B^0_d\to {D}^{\prime -}_{1\ (1/2)} \ell \nu) &\simeq& 0.7 \times  10^{-3}\,, \nn\\
{\cal B}(B^0_d\to {D}_0^{\ast -} \ell \nu) &\simeq& 0.6 \times 10^{-3}\,.
\end{eqnarray}
Finite width effects are not negligible in the case of broad states, but they would reduce the predictions (by about $20 \%$), thus further aggravating the problem we are addressing, i.e. the problem that  predictions seem to be too small with respect to experiment.
\vskip 0.5cm
\subsubsection{Non leptonic $B \rightarrow \overline{D}^{\ast\ast}\pi^+$ decays in theory} \label{NL}
\begin{figure}[t!]
\begin{center}
{\resizebox{7.7cm}{!}{\includegraphics{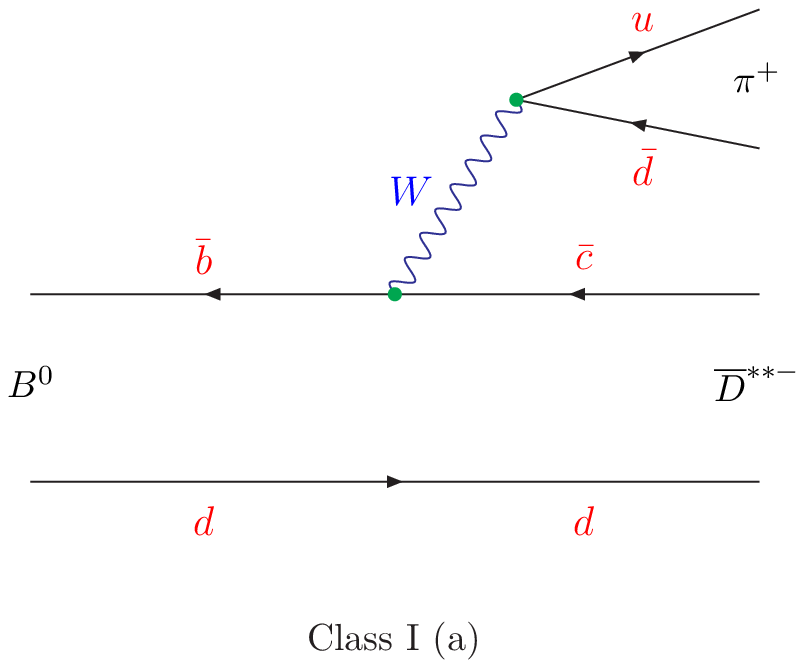}}}\qquad 
{\resizebox{7.7cm}{!}{\includegraphics{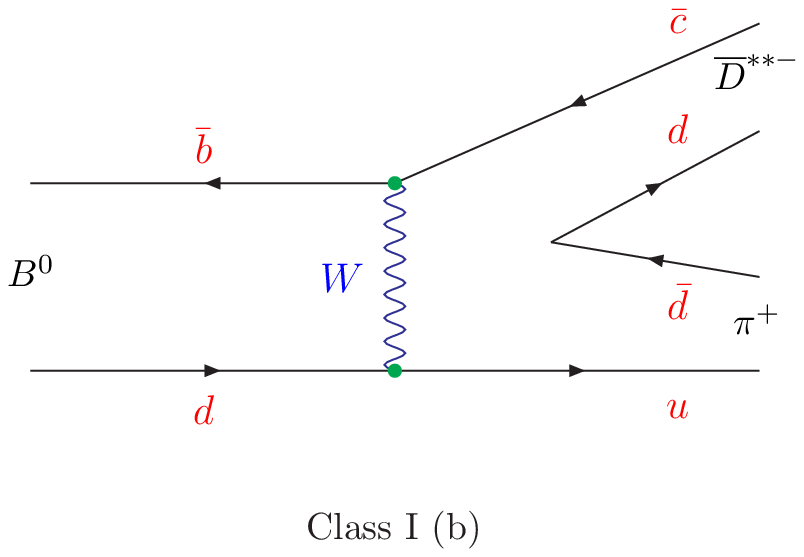}}} 
\caption{\label{fig:1}\footnotesize{ Diagrams contributing to the Class I non-leptonic decay $B\to \overline{D}^{\ast\ast}\pi$: (a) pion emission, (b) weak annihilation. }} 
\end{center}
\end{figure}
Semileptonic decays would  in principle provide the cleanest test of the theoretical predictions, but the undetected neutrino prevents us from doing a very good analysis. The above predictions can fortunately be  tested by considering the non-leptonic decays, if an extra assumption is made, namely factorization. 
\begin{figure}[h!]
\begin{center}
{\resizebox{7.7cm}{!}{\includegraphics{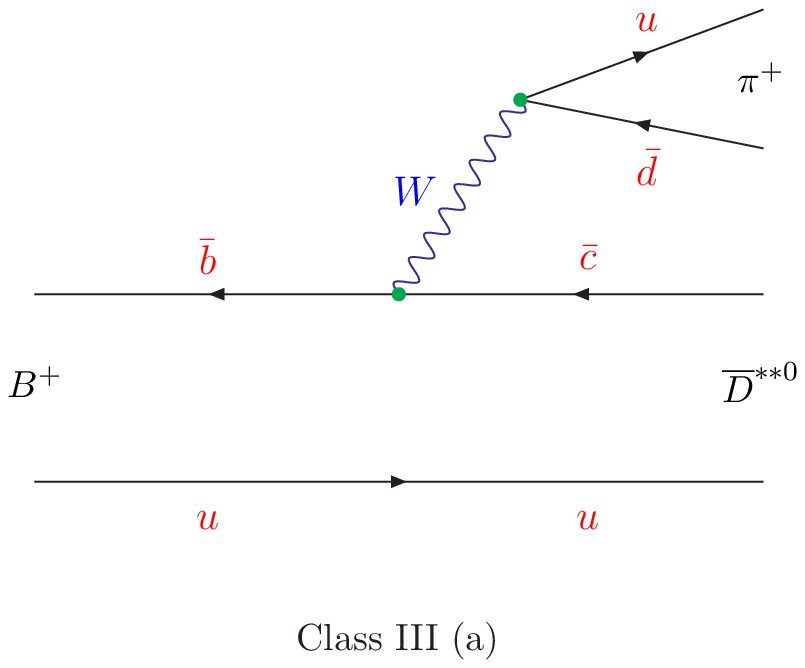}}}\qquad 
{\resizebox{7.7cm}{!}{\includegraphics{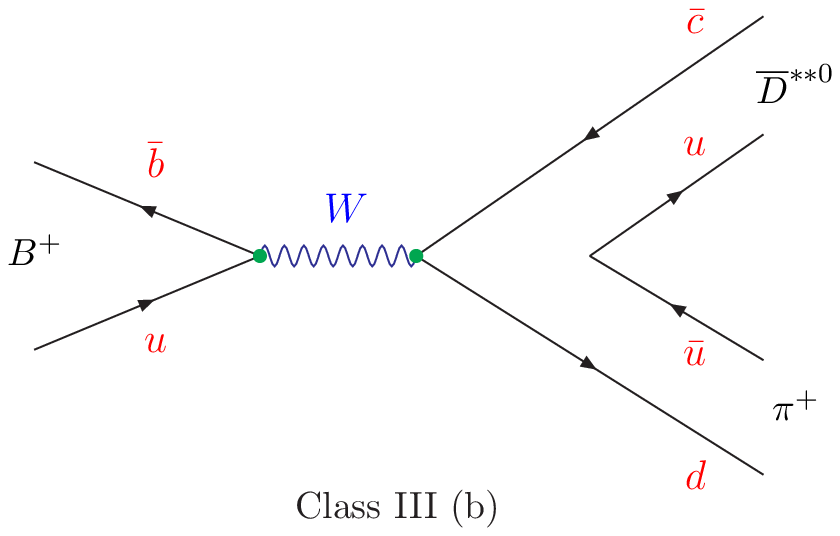}}}\\ 
\hfill \\
{\resizebox{7.7cm}{!}{\includegraphics{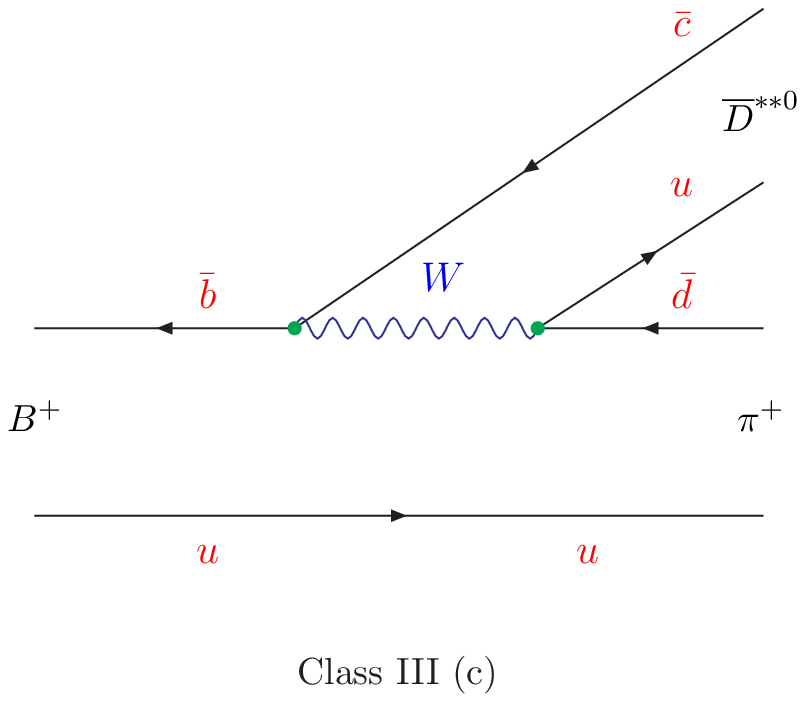}}} 
\caption{\label{fig:2}\footnotesize{ Diagrams contributing to the Class III non-leptonic decay $B\to \overline{D}^{\ast\ast}\pi$: (a) pion emission, (b) weak annihilation, (c) $\overline{D}^{\ast\ast}$-meson emission.}} 
\end{center}
\end{figure}

As it is well known, there are  three classes of non-leptonic decays. $B_d^0 \rightarrow \overline{D}^{(\ast,\ast\ast)-} \pi^+$, for example, belongs to Class I and is described
by the sum of two diagrams:  the pion emission through $W$ (which is colour favored, c.f. fig.~\ref{fig:1}a), and the annihilation through the $W$ exchange (shown in fig.~\ref{fig:1}b). The annihilation is expected to be small and the pion emission can be easily evaluated in the factorization approximation as a product of the annihilation constant $f_{\pi}$ and the $B \to  \overline{D}^{\ast\ast}$ form factor. As before, we use the form factors computed in the heavy quark limit, whereas in the phase space computation we use the physical meson masses. Using the values given in eq.~(\ref{eq:BTGI}) one has~\cite{jugeau}:~\footnote{The expressions for the amplitudes involve the coefficient $a_1$~\cite{stech}, for which we take $a_1 \simeq 1$.}
\begin{eqnarray}\label{NL0plus}
{\cal B}(B^0_d\to{D}_2^{\ast -} \pi^+) &\simeq& 1.1 \times 10^{-3} \,,\nn\\
{\cal B}(B^0_d\to {D}^{-}_{1\ (3/2)}\pi^+ )&\simeq& 1.3 \times  10^{-3} \,,\nn \\
{\cal B}(B^0_d\to {D}^{\prime -}_{1\ (1/2)} \pi^+ ) &\simeq& 1.1 \times  10^{-4}\,, \nn\\
{\cal B}(B^0_d\to {D}_0^{\ast -}  \pi^+ ) &\simeq& 1.3 \times 10^{-4}\,,
\end{eqnarray}
where we include the $w$-dependence of $\tau_{1/2,3/2}(w)$, away from $w=1$, which reduces the rate by around a factor of $2$.
The qualitative picture one gets from this exercise is that, similarly to the case of semileptonic decays, the decay rates to $j=1/2$ states should be an order of magnitude smaller with respect to those with $j=3/2$ in the final state.

If one considers a class III decay, such as $B^+ \rightarrow \overline{D}^{(\ast,\ast\ast)0} \pi^+$,  
then a priori three diagrams contribute:  (i) pion emission through a color suppressed $W$-exchange (see fig.~\ref{fig:2}a), (ii) annihilation of $B$ through $W$, shown in fig.~\ref{fig:2}b, which is negligible because of the factor $\propto V_{ub}$, (iii)  emission of the $\overline{D}^{\ast\ast}$ meson  through $W$-exchange (see fig.~\ref{fig:2}c).  Although color suppressed, the last diagram cannot be neglected for the decay to $j=1/2$ because its size is similar to the pion emission. This is a consequence of the smallness of $\tau_{1/2}(w)$~\cite{jugeau}. On the other hand it vanishes for $D_2^{\ast}$, where the factor $f_{D_2^{\ast}}$, that appears in the factorized expression of the amplitude, vanishes because the $2^+$ state does not  couple to the weak current. Notice that $f_{D_{1}} \equiv f_{D_1^{3/2}}$ is also expected to be small, based on the heavy quark symmetry. 

Since there is only one sizable contribution, class I decays should be preferred to test the theoretical estimate of the $B \rightarrow \overline{D}^{(\ast,\ast\ast)}$ form factors. Class III nevertheless offers an additional qualitative test, because the additional diagram leads to a large difference with class I.  In principle, the cleanest way to assess the magnitude of $\tau_{1/2}(w)$ and $\tau_{1/2}(w)$ would be through the study of semileptonic $B\to D^{\ast\ast}$ decays, because the experimental extraction from the class I nonleptonic decays could be spoiled by the presence of the resonant $\pi\pi$ pair in the final state.

\section{Experimental situation}
\label{sec:experimental}\vskip 0.5cm
In contrast to the consistency of theoretical approaches, we find a rather different situation on the experimental side,
especially in semileptonic decays where blatant inconsistencies are found between two sets of measurements by BaBar and Belle. We begin by explaining why 
semileptonic decays are in principle more difficult to analyze than non-leptonic three-body decays like 
$B \to \overline{D}^{(\ast)} \pi \pi$, which may seem paradoxical since the former have a much larger rate.

Dalitz plot analyses of $B \rightarrow \overline{D}^{(\ast)}\pi\pi$ decay channels at B-factories have provided informations on the production rate and on the resonance parameters of broad $D^{\ast\ast}$ resonances. Events are selected if the energy of the candidate is compatible with the beam energy and if the mass of the system formed by its decay products is compatible with the nominal B-meson mass. For an integrated luminosity of 500 fb$^{-1}$ and an assumed decay branching fraction of $10^{-3}$ there are typically $4000$ and $9000$ reconstructed signal events for the
$D^{\ast\ast} \rightarrow D^{\ast +}\pi^-,~D^{\ast +}\rightarrow D^0 \pi^+,~D^0 \rightarrow K^-\pi^+,~K^-\pi^+\pi^+\pi^-$ and
$D^{\ast\ast}\rightarrow D^+ \pi^-,~D^+\rightarrow K^- \pi^+ \pi^+$ decay chains, respectively.

Because of the missing neutrino, $B$-meson semileptonic decays are more difficult to analyze. It is necessary to fully  reconstruct the other $B$-meson ($B_{tag.}$) and a cut on the missing mass squared is used
to select events with only a missing neutrino (Belle) or keeping also events in which the soft pion from the cascade $D^* \rightarrow D \pi$
escapes detection, in addition to the neutrino (BaBar). These analyses have an efficiency which is typically two orders of magnitude lower than for the exclusive $B \rightarrow 3$-body decays considered
previously. In practice, because semileptonic branching fractions into
individual $D^{\ast\ast}$ states are an order of magnitude higher than in exclusive non-leptonic final states, there is typically only an order of magnitude difference between the statistics of signal events analyzed in
non leptonic and semileptonic $B$-meson decays.

\subsection{A long-standing confusion in semileptonic decays}

Unless explicitly stated, the numbers presented in this section are obtained by using the values given
by the HFAG collaboration~\cite{ref:hfag}, and we average measurements from neutral and charged $B$-meson using isospin symmetry.
Obtained values are quoted for the $B_d^0$ meson.

The inclusive semileptonic decay branching fraction of $B_d^0$ and of 
$B^+$ decay is far from being saturated by the sum of $\overline{D} \ell^+ \nu_{\ell}$
and $\overline{D}^\ast \ell^+ \nu_{\ell}$ decay channels. More specifically~\cite{ref:bellebsl}:
\begin{eqnarray}
{\cal B}(B_d^0 \rightarrow \overline{X}_c \ell^+ \nu_{\ell}) & =& (10.09 \pm 0.22)\%\,, \\
{\cal B}(B_d^0 \rightarrow \overline{D} \ell^+ \nu_{\ell}) & =& (2.12 \pm 0.06)\%\,,\nonumber\\
{\cal B}(B_d^0 \rightarrow \overline{D}^* \ell^+ \nu_{\ell}) & =& (5.11 \pm 0.10)\%\,.\nonumber
\end{eqnarray}
In other words, the semileptonic branching fraction to the charm states which are not simply a $D$
or a $D^\ast$ is thus equal to:
\begin{equation}
{\cal B}(B_d^0 \rightarrow {\rm non-}\overline{D}^{(*)} \ell^+ \nu_{\ell})=
(2.86 \pm 0.25)\%. 
\end{equation}
Decays to the narrow $D^{\ast\ast}$ states have been measured with good accuracy:
\begin{eqnarray}
{\cal B}(B_d^0 \rightarrow \overline{D}_2^* \ell^+ \nu_{\ell}) & =& (0.29 \pm 0.03)\%\,,\nonumber \\
{\cal B}(B_d^0 \rightarrow \overline{D}_1 \ell^+ \nu_{\ell}) & =& (0.58 \pm 0.05)\%\,,
\end{eqnarray}
giving:
\begin{equation}
{\cal B}(B_d^0 \rightarrow \overline{D}^{\ast\ast}_{\rm narrow} \ell^+ \nu_{\ell})=
(0.87 \pm 0.06)\%\,. 
\end{equation}

The above values include the branching fraction of ${D}^{\ast\ast}$ into the observed final state.~\footnote{
Few branching fractions of ${D}^{\ast\ast}$ decays into exclusive final states are not well determined and we use the following values:
\begin{align}
&{\cal B}(D_2^{*0} \rightarrow D^+ \pi^-) = 0.41 \pm 0.02\,,\qquad {\cal B}(D_2^{*0} \rightarrow D^{*+} \pi^-) = 0.26 \pm 0.02\,,\qquad {\cal B}(D_0^{*0} \rightarrow D^+ \pi^-) = 2/3 \,,\nonumber\\
&{\cal B}(D_1^{0} \rightarrow D^{*+} \pi^-) = 0.45 \pm 0.02\,,\qquad {\cal B}(D_1^{0\prime} \rightarrow D^{*+} \pi^-) = 2/3\,.
\end{align} 
To make these evaluations we have assumed  in addition that:
(i) $D_2^*$ decays exclusively into $D\pi$ or $D^*\pi$.
Channels with two charged pions have been studied and no signal was observed; 
(ii) $D_0^*$ decays exclusively into $D\pi$;
(iii)  $D_1$ decays into $D\pi\pi$ and $D^*\pi$ with a ratio  ${{\cal B}(D_1^{0} \rightarrow D^0 \pi^+ \pi^-)/{\cal B}(D_1^{0} \rightarrow D^{*+}\pi^-)} = 0.32 \pm 0.03$, and we assume
that the decay proceeds through the chain $D_1 \rightarrow D_0^{*}\pi$.
}
Another piece of information comes from the measurements of
the exclusive $B \rightarrow \overline{D}^{(\ast)}\pi\ell^+ \nu_{\ell} $ decays:
\begin{eqnarray}
{\cal B}(B_d^0 \rightarrow \overline{D} \pi \ell^+ \nu_{\ell}) & =& (0.60 \pm 0.06)\%\,,\nonumber\\
{\cal B}(B_d^0 \rightarrow \overline{D}^\ast \pi \ell^+ \nu_{\ell}) & =& (0.83 \pm 0.06)\%\,, 
\end{eqnarray}
giving:
\begin{equation}
{\cal B}(B_d^0 \rightarrow \overline{D}^{(\ast)}\pi\ell^+ \nu_{\ell})=
(1.43 \pm 0.08)\%. 
\end{equation}
This value can be compared with the expected $\overline{D}^{(\ast)}\pi\ell^+ \nu_{\ell}$ 
and $\overline{D}\pi\pi\ell^+ \nu_{\ell}$
branching fractions from the decays of narrow $\overline{D}^{\ast\ast}$ states
given in Table \ref{tab:narrow_decays}.

\begin{table}[!h]
\begin{center}
  \begin{tabular}{|c|c|c|c|}
    \hline
 Decay channel  & $\overline{D}_1$  & $\overline{D}_2^*$ & total \\
\hline
 ${\cal B}(B_d^0 \rightarrow \overline{D}\pi \ell^+ \nu_{\ell})$& $-$ & $(0.18 \pm 0.02)\%$ & $(0.18 \pm 0.02)\%$\\
\hline
 ${\cal B}(B_d^0 \rightarrow \overline{D}^*\pi \ell^+ \nu_{\ell})$& $(0.39 \pm 0.04)\%$ & $(0.11 \pm 0.01)\%$& $(0.50 \pm 0.04)\%$\\
\hline
 ${\cal B}(B_d^0 \rightarrow \overline{D}\pi \pi\ell^+ \nu_{\ell})$& $(0.19 \pm 0.02)\%$ & $0.00$& $(0.19 \pm 0.02)\%$\\

\hline
  \end{tabular}
  \caption[]{\small {Branching fractions for 
$B_d^0 \rightarrow \overline{D}^{(*)} \pi (\pi)\ell^+ \nu_{\ell} $ decay channels
where the hadrons cascade from a narrow $\overline{D}^{\ast\ast}$ meson.}
  \label{tab:narrow_decays}}
\end{center}
\end{table}

From these measurements one can draw several conclusions:
\begin{itemize}
\item narrow $\overline{D}_1$ and $\overline{D}_2^*$ states, 
with no additional pion,
account for about 1/3 of the 
non-$\overline{D}^{(*)}\ell^+ \nu_{\ell}$ final states;
\item $\overline{D}^{(*)}\pi \ell^+ \nu_{\ell}$ final states account for about 1/2
of the non-$\overline{D}^{(*)}\ell^+ \nu_{\ell}$ final states. As a result,
final states with two or more pions should account for the other half;
\item the broad state
component of the $\overline{D}\pi$ system corresponds to a branching fraction equal to
${\cal B}(B_d^0 \rightarrow [\overline{D}\pi]_{broad} \ell^+ \nu_{\ell})=(0.42 \pm 0.06)\%$;
\item the broad state
component of the $\overline{D}^*\pi$ system corresponds to a branching fraction equal to
${\cal B}(B_d^0 \rightarrow [\overline{D}^*\pi]_{broad} \ell^+ \nu_{\ell})=(0.33 \pm 0.07)\%$;
\end{itemize}

For theorists, it remains to interpret the origin of the 
broad $\overline{D}^{(\ast)}n\pi$ components, with $n \geq 1$ which correspond to 2/3 
of these hadronic final states in $B$ semileptonic decays.

There is, at present, an apparent contradiction between the measured
values for the $\overline{D}_0^*$. 
\begin{eqnarray}
{\cal B}(B_d^0 \rightarrow \overline{D}_0^* \ell^+ \nu_{\ell}) & =& (0.35 \pm 0.07)\%\,,
\end{eqnarray}
and the corresponding theoretical
expectations.
According to theory, the production of these broad resonances
should be much lower than the one of narrow states and this is apparently
not verified (see below for details).
For the $\mathrm{  broad} ~\overline{D}_1^{\prime}$ state, the situation is different because
the two experiments  disagree. Belle does not see any
$\mathrm{ broad}~\overline{D}_1^{\prime}$
component,
while BaBar gives :
\begin{equation}
{\cal B}(B_d^0 \rightarrow \overline{D}_1^{\prime}\ell^+ \nu_{\ell})  = (0.26 \pm 0.04 \pm 0.04) \%\,,\nonumber                                               \end{equation} 
HFAG gives (``Updates of Semileptonic Results for End Of 2009") 
\begin{equation}
{\cal B}(B_d^0 \rightarrow \overline{D}_1^{\prime}\ell^+ \nu_{\ell})  = (0.13 \pm 0.06)\%\,,\nonumber                                               
\end{equation}
but it must be understood that the two measurements are incompatible  
(BaBar and Belle results differ by $3.2~\sigma$.)
The PDG group discards Belle without explanation.

Meanwhile several comments are in order:

\begin{itemize}
\item experimenters cannot claim that they have really measured
the production of the broad $\overline{D}_0^*$ and  $\overline{D}_1^{\prime}$ resonances. There could
be additional contributions from broad $\overline{D}\pi$ and $\overline{D}^*\pi$ final states
in the registered spectra; BaBar states explicitly that they have 
not subtracted any non resonant background, for lack of a satisfactory fit for it;

\item the branching fraction attributed to the $\overline{D}_0^*$ is compatible
with the broad component rate obtained by analyzing the $\overline{D}\pi$ final state;

\item 
for the $\overline{D}_1^{\prime}$ production, the quoted value of BaBar
is compatible with the broad component rate obtained by analyzing the 
$\overline{D}^*\pi$ final state.
\end{itemize}

All this is compatible with the idea that the real difficulty causing the disagreement within experiment, and perhaps
with theory is the one of {\it analyzing events in terms of broad resonances}, as we discuss in subsection \ref{broad}.

\subsubsection{Summary}
In Table \ref{tab:bsl} are summarized the present measurements
of $B$-meson semileptonic decays into a charm hadronic system. Values are 
quoted for the $B_d^0$; corresponding results for the $B^+$ can be
obtained by multiplying these values by the lifetime ratio
$\tau(B^{+})/\tau(B_d^{0})=1.079\pm 0.007$. 

\begin{table}[!htb]
\begin{center}
  \begin{tabular}{|c|c|}
    \hline
 Decay channel  & branching fraction ($\%$)\\
\hline
 $B_d^{0} \rightarrow \overline{X}_c \ell^+ \nu_{\ell}$& $10.09 \pm 0.22 $\\
 $B_d^{0} \rightarrow D^- \ell^+ \nu_{\ell}$& $2.12 \pm 0.06 $\\
 $B_d^{0} \rightarrow D^{*-} \ell^+ \nu_{\ell}$& $5.11 \pm 0.10 $\\
 $B_d^{0} \rightarrow D_1^{-} \ell^+ \nu_{\ell}$& $0.58 \pm 0.05 $\\
 $B_d^{0} \rightarrow D_2^{*-} \ell^+ \nu_{\ell}$& $0.29 \pm 0.03 $\\
 $B_d^{0} \rightarrow [\overline{D} \pi]_{broad} \ell^+ \nu_{\ell}$& $0.42 \pm 0.06 $\\
 $B_d^{0} \rightarrow [\overline{D}^* \pi]_{broad} \ell^+ \nu_{\ell}$& $0.33 \pm 0.07 $\\
 $B_d^{0} \rightarrow [\overline{D} \pi\pi]_{narrow} \ell^+ \nu_{\ell}$& $0.19 \pm 0.02 $\\
 $B_d^{0} \rightarrow \overline{X}_{c,~broad}^{remaining} \ell^+ \nu_{\ell}$& $1.24 \pm 0.26 $\\
 $B_d^{0} \rightarrow D_s^{(*)-} K^0\ell^+ \nu_{\ell}$& $0.06 \pm 0.01 $ \cite{ref:babardsk}\\
\hline
  \end{tabular}
  \caption[]{\small {Measured semileptonic $B_d^{0}$ branching fractions.
The $[\overline{D} \pi\pi]_{\rm narrow}$ hadronic final state corresponds
to the decay of the $D_1^{-}$. The $\overline{X}_{c,\rm broad}^{\rm remaining}$
hadronic final state contains a $\overline{D}$ or $\overline{D}^*$ meson with at least
two pions or a $\eta$ or a $\eta^{\prime}$ meson.}
  \label{tab:bsl}}
\end{center}
\end{table}
From these measurements there are at least two questions which remain to
be clarified:
\begin{itemize}
\item the 
origin
of $[\overline{D}^{(*)} \pi]_{broad}$ states. What is
the fraction of these states which can come from the $\overline{D}_{0}^*$ 
and $\overline{D}_1^{\prime}$ mesons? A possible answer to this question
is the subject of the present paper.

\item the contribution of broad final states with several pions or with
a $\eta$ or a $\eta^{\prime}$. Because of the large mass of the
$\eta^{(\prime)}$ mesons it is not expected that corresponding final states
have a large contribution.

\end{itemize}

\subsection{$B \rightarrow \overline{D}^{\ast\ast}\pi^+$ decays}

In this subsection we provide a summary of present measurements at BaBar
and Belle of the decays $B \rightarrow \overline{D}^{\ast\ast}\pi^+$.

BaBar and Belle collaborations have measured several
$B \rightarrow \overline{D}^{\ast\ast}\pi^+$ decay channels using Dalitz
analyses. 
Averaged values of $B \to \overline D^{\ast\ast} \pi$ branching fractions measured by 
BaBar~\cite{ref:babar_dpipich, ref:babar_dpipine}
and Belle~\cite{ref:belle_dpipich, ref:belle_dpipine1, ref:belle_dpipine2} are given in Table \ref{tab:dsstarpi}.

\begin{table}[!htb]
\begin{center}
  \begin{tabular}{|c|c|c|}
    \hline
 Decay channel  & $B_d^{0}$  & $B^{+}$ \\
\hline
 $\overline{D}_2^* \pi^+$& $(4.9 \pm 0.7)\times 10^{-4}$& $(8.2 \pm 1.1)\times 10^{-4}$ \\
\hline
 $\overline{D}_1 \pi^+$& $(8.2^{+2.5}_{-1.7})\times 10^{-4} $& $(15.1 \pm 3.4)\times 10^{-4}$ \\
\hline
 $\overline{D}_1^{\prime} \pi^+$& $<1 \times 10^{-4}$ & $(7.5 \pm 1.7)\times 10^{-4}$ \\
\hline
 $\overline{D}_0^* \pi^+$& $(1.0 \pm 0.5)\times 10^{-4}$& $(9.6 \pm 2.7)\times 10^{-4}$\\
\hline
  \end{tabular}
  \caption{\small {Measured branching fractions for 
$B \rightarrow \overline{D}^{**} \pi^+$ decay channels.}
  \label{tab:dsstarpi}}
\end{center}
\end{table}
A few remarks can be made:
\begin{itemize}
\item branching fractions are higher for the $B^+$ than for the $B_d^0$, where both are measured.~\footnote{In these comparisons between branching fractions for charged and neutral $B$-mesons we are interested in differences which appear in addition to the $7$~\% expected from the lifetime difference.}
\item considering the $\overline{D}_2^*$ production, which is the most
accurate,
it is also  not  too far from equality, as would be expected according to factorization, since there is no diagram with $\overline{D}_2^*$ emission.
On the contrary, it is expected that for the $0^{+}$ the two rates should be very different, as it is indeed found (see below).

\item $\overline{D}_1$ production seems to be higher than $\overline{D}_2^*$,
 in a certain contradiction with heavy quark symmetry. This is understandable by a simple $1/m_c$ effect, as in semileptonic decays.
\item the production of $\overline{D}_0^*$ states is not well measured. In $B^+$
decays it seems to be similar to the $\overline{D}_2^*$ but in $B_d^0$ decays it 
seems to be much smaller. In fact, measurements of $B_d^0$ decays
from Belle and BaBar (preliminary) are not in good agreement even if it
is difficult to draw a clear conclusion because of the attached uncertainties:
\begin{align}   
{\cal B}(B_d^0 \rightarrow D_0^{*-} \pi^+)&\times~{\cal B}(D_0^{*-}\rightarrow \overline{D}^0 \pi^-)=\\
&(0.60\pm0.13\pm0.15\pm0.22)\times 10^{-4} ~{\rm Belle}\,,\nonumber\\
 &  (2.18\pm0.23\pm0.33\pm1.15\pm0.03) \times  10^{-4}~{\rm BaBar}\,.\nonumber
\end{align}
BaBar reports a larger systematic uncertainty, coming from the modelling 
of the fitted distribution, than Belle.
Anyway, the decay of neutral $B_d^0$ is in both experiments clearly smaller than the charged one,
and this can be understood theoretically because in the charged case, and, contrarily to $\overline{D}_2^{**,0}$, there is a diagram with emission of $\overline{D}_0^{**,0}$ which can overwhelm the pion emission diagram, which is small because of the smallness of $\tau_{1/2}(1)$.
\end{itemize}

\vskip 0.5cm
\section{Comparison between theory and experiment}
\label{sec:comparison}

Results of the preceding sections are summarized in Table \ref{tab:theoexpt}. Let us then recapitulate the conclusion one can draw by taking the experimental data as they are presented.

\begin{table}[!htbp!]
\begin{center}
{\small
  \begin{tabular}{|c|c|c|c|}
    \hline
 &${\cal B}_{theory}$  & ${\cal B}_{expt.}$ & ${\cal B}_{expt.}/{\cal B}_{theory}$\\
\hline
$B^0_d \rightarrow \overline{D}^{\ast\ast}e^+\nu_e$ & & &\\
\hline
$\overline{D}_2^*$ & $0.7 \times 10^{-2}$ & $(0.29 \pm 0.03) \times 10^{-2}$ & $\sim 0.5$\\
$\overline{D}_1$ & $0.45 \times 10^{-2}$ & $(0.58 \pm 0.05) \times 10^{-2}$ & $\sim 1.$\\
$\overline{D}_1^{\prime}$ & $0.7 \times 10^{-3}$ & $[0.,3.2] \times 10^{-3}$ & $[0., 5.]$\\
$\overline{D}_0^*$ & $0.6 \times 10^{-3}$ & $(3.5 \pm 0.7) \times 10^{-3}$ & $6. \pm 1.$\\
\hline
$B^0_d \rightarrow \overline{D}^{\ast\ast}\pi^+$ & & &\\
\hline
$\overline{D}_2^*$ & $1.1 \times 10^{-3}$ & $(0.49 \pm 0.07) \times 10^{-3}$ & $\sim 0.5$\\
$\overline{D}_1$ & $1.3 \times 10^{-3}$ & $(8.2 ^{+2.5}_{-1.7})\times 10^{-4}$& $[0.5,1.]$\\
$\overline{D}_1^{\prime}$ & $1.1 \times 10^{-4}$ & $<10^{-4} (90\%~ C.L.)$ & no result\\
$\overline{D}_0^*$ & $1.3 \times 10^{-4}$ & $[0.3,3.4]\times 10^{-4}$ & $[0.2,2.6]$\\
\hline
  \end{tabular}
}
  \caption[]{\small {In this Table are collected the values expected and measured
for $\overline{D}^{\ast\ast}$ production in semileptonic and non leptonic $B^0_d$ meson decays.
These values have been given already in previous sections. The theoretical expectation is taken to be the one of the quark model, subsection \ref{phenomenological}. A range of values
is given within brackets when there is not a good compatibility between BaBar and Belle measurements. In this case we take the minimum value minus one sigma
and the maximum value plus one sigma to define this range. In general there is
agreement between measured and expected branching fractions for narrow states.
For broad states results are in contradiction with expectations (mainly
the $\overline{D}_0^*$ production in semileptonic decays) or rather uncertain.}
  \label{tab:theoexpt}}
\end{center}
\end{table}

\subsection{Ratio of $B^0_d \to D_0^{*-} \ell^+ \nu_{\ell}$ and  $B^0_d  \to D_0^{*-}  \pi^+$ }

Assuming the validity of the QCD factorization and by describing the $B\to \overline{D}^{\ast\ast}$ transition matrix elements by a slowly varying $\tau_{1/2,3/2}(w)$, one can easily see that
$B^0_d  \to D_0^{*-}  \pi^+$ and $B^0_d \to D_0^{*-} \ell^+ \nu_{\ell}$ decays
are governed by $\tau_{1/2}$ alone.~\footnote{The general idea of the relation between semileptonic and non leptonic decays is due to M.~Neubert~\cite{neubert}.} 
Using the values given in Table \ref{tab:theoexpt}, the ratio of semileptonic to non-leptonic decays with $\overline{D}^\ast_0$ in the final state must be
$\simeq 5$. Experimentally, instead, such a ratio spans a large interval between 8 and 140.
In contrast to that situation, decays to the narrow $\overline{D}_2^\ast$ state 
lead to a ratio that is theoretically expected to be equal to 6, which is confirmed by the experimentally established value $6 \pm 1$.  
For decays to $\overline{D}_1$ state, uncertainties are larger and based
on a single unpublished result from Belle but the expected theoretical value
for the ratio, which is equal to 3.5, agrees roughly with experiment ($7 \pm 2$).

\subsection{Contradiction between the phenomenological predictions and the semileptonic experimental data }

Now, we can go further still and state that  the {\sl semileptonic} experimental data contradicts the HQET estimate for the decay to a $j=1/2$ state, with a huge discrepancy which is one order of magnitude in rate.  

To arrive to such a conclusion, one first has to take into account the disagreement among experiments in ${\cal B}(B\to \overline{D}_1^\prime \ell^+ \nu_{\ell})$ states. While the result reported by Belle seems to be compatible with expectation of a very small rate, the result of BaBar is much larger and disagrees with both Belle and the expected value. Both experiments, instead, agree on the value for ${\cal B}(B\to \overline{D}^\ast_0 \ell^+ \nu_{\ell})$  which is far too large when compared with expectations.  While the results by BaBar are far too large when compared to the expectations, they are still consistent with the heavy quark symmetry expectations, i.e. the two rates are nearly equal. The results by Belle instead indicate a complete breakdown of the heavy quark symmetry. On the whole, it is fair to say that both experiments disagree with theory for both $j=1/2$ states.

On the other hand, there is a qualitative agreement in both types of transitions to $j=3/2$ states. There is an excess of theory, by a factor two, for ${\cal B}(B\to \overline{D}_2^\ast \ell^+ \nu_{\ell})$, but there is also an overall success for the sum ${\cal B}(B\to [\overline{D}_2^\ast, \overline{D}_1] \ell^+\nu_{\ell}) \simeq 1\% $.

\subsection{Better situation for non leptonic decays, yet not conclusive}

The situation with non-leptonic decay to a $j=1/2$ state is much better not only in experiment, but also concerning the comparison between theory and experiment. For the Class I decay, $B^0_d \rightarrow D_0^{*-}\pi^+$,  the prediction~(\ref{NL0plus}) coincides with the Belle measurement, and is compatible with BaBar 
within the quoted uncertainties. Notice the important point that in Class I decays factorization is expected to hold to a good approximation both on theoretical grounds and also, taking into account a large number of decays with such topology, on empirical grounds.

The discrepancy between Belle and BaBar occurs in  
$B^0_d \rightarrow D_0^{*-}\pi^+$, which could be attributed to the difficulty of extracting a broad resonance, with possible large non-resonant structure, and with the additional difficulty of a $\pi \pi$ crossed channel interference (see below). 

A fact that seems to attest the soundness of the theoretical statements about the smallness of the production of the $j=1/2$ states is the large difference between neutral and charged $B$-decay into 
the broad $\overline{D}_0^\ast$-state: the charged decay rate is much larger than the neutral one, by about one order of magnitude as given in 
Table \ref{tab:dsstarpi}. 
This is easily understood 
because
an additional diagram is present in the charged case, which is the $\overline{D}^{\ast\ast}$-emission (class III).~\footnote{Such an explanation was first offered by Belle \cite{abeB0}.}  Although color suppressed, this diagram gives a contribution much larger than the one with the pion emission, if $\tau_{1/2}(1)$ is small~\cite{jugeau}. In that case the $\overline{D}^{\ast\ast}$ emission amplitude dominates the charged rate, and dominates over the neutral decay amplitude. A similar effect is observed in the case of the broad $1^+$ final meson. Although a branching ratio has not been published, the bound on the neutral $B$-decay in ref.~\cite{abeB0} clearly indicates that the charged decay is much larger than the neutral one.

The discrepancy of around a factor of two between charged and neutral $B$ decay to  $\overline{D}_2^\ast$, could be interpreted as an estimate of the correction to the factorization approximation in which the two decays are expected to have nearly equal rates. Such a discrepancy is similar to what is found in common tests of factorization~\cite{jugeau,stech,otherfact1, otherfact2, otherfact3} (see also references therein).

\subsection{Discussion of the main discrepancy and  possible explanations}

If we believe the results of theory, which are rather consistent, and if we take the experimental results for broad states in semileptonic decays,  then one or both states have much too large rates as compared to theory. One experiment also suggests a complete breaking of heavy quark symmetry. 
In non-leptonic decays there is a better agreement between theory and experiment
but present uncertainties in $B^0_d \rightarrow \overline{D}^{\ast\ast}\pi$ decays
are too large to derive firm conclusions.

Of course, one could evoke weaknesses in the assumptions which allow to derive {\sl phenomenological} predictions. In particular one can argue that the  $1/m_c$ effects could be large. However, large $1/m_c$ effects cannot explain the contrast between a relative success in non-leptonic decays where they should be present too. One could also complain about he validity of the  factorization approximation, but that is unlikely to be the case as factorization in the Class I decays passed many experimental tests and no large deviations have been found so far. 
Finally, let us stress the satisfactory qualitative agreement in the case of decay to a $j=3/2$ state, both semileptonic and non-leptonic ones.

\underline{\bf The problem of broad resonances} \label{broad}. A possible reason for the qualitative agreement between theory and experiment in the $B$-decays to a $j=3/2$ state can be explained by the fact that the $j=3/2$ states are narrow. Distinguishing very broad resonances from continuum is extremely difficult enterprise, both on theoretical and experimental sides.  

There is no unambiguous way of writing the broad resonance line shape, all the more for $S$-wave scattering where very strong couplings can be present, and therefore the very notion of separating a resonance and the non-resonant continuum is theoretically  ambiguous.  Furthermore, the $q \bar q$ states could be competing with non $q \bar q$ states in $S$ waves and additional resonances could be generated by the scattering. Finally one can also encounter problems with contributions arising from the tails of the ground state (denoted as $\overline{D}_v^*, B_v^*$ in tab.~II of ref.~\cite{ref:belle_dpipine2}) or of radial excitations in $\overline{D}^{(*)} \pi$ .

Ideally, one should be able to compare the whole amplitude with experiment, and not just the resonance under study, but that is obviously not possible in practice. All this underlines the advantage of working with narrow resonances.

It must be repeated, however, that if broadness was a sole cause for a large discrepancy discussed above, then one would be short of explanation regarding the non-leptonic decays for which the disagreement is not large. 
 Keep in mind, however, that potentially large uncertainties due to the arbitrariness of the non-resonant continuum should enter the game also in the non-leptonic case. Last but not least, for neutral $ B^0_d\to D_0^{\ast -} \pi^+ \to \overline{D}^0 \pi^-\pi^+$, which is the relevant one for our purpose, one can have interference with the crossed channel $\pi \pi$ which resonates into $\rho, f_0, etc ...$ ($ B^0_d\to \overline{D}^0 \rho^0$, ...) All these contributions cannot be separated out without heavily relying on specific models and the resulting uncertainty may lead to inconclusive comparison between theory and experiment. 
 
Blaming broadness of states for the difficulties in measuring the rates of $j=1/2$ is strongly supported by the following argument. In $B^0_d \to \overline{D}^0 \pi^- \pi^+$, Belle and BaBar find exactly the same total rate, and 
the same rate for all the decays to relatively {\sl narrow} resonances, i.e. not only $B \to \overline{D}_2^* \pi$, but also $B \to \overline{D} \rho$, $B \to \overline{D} f_2(1235)$. On the other hand, large discrepancies appear in the central values of the decays to broad resonances, not only in $B \to \overline{D}_0^* \pi$, but also in $B \to \overline{D} f_0(600)$ ($S$-wave).~\footnote{Note that PDG use the notation $f_0(600)$ for the lowest scalar $J^{PC}=0^{++}$ state~\cite{ref:pdg}, that is often referred to as $\sigma(600)$, or $\epsilon(600)$.}

\vskip 0.5cm
\part{\underline{\Large Proposal for the complementary study of the narrow}  \\ 
\underline{\Large strange counterparts}
\label{sec:proposal}}

\section{Motivation}
\label{sec:motivation}

Our proposal starts from the above observation that analysis of \underline{broad resonances} has always been a difficult task. The fact that no special problem arises for the narrow $j=3/2$ states suggests that  the broadness of $j=1/2$ states in the non-strange case could be the origin of the difficulties. At least, it could help much if one could deal with states analogous to the controversial $D^{\ast\ast}$ (i.e. $D_0^\ast$ and $D_1^\prime$), but narrow. Even if not leading to an immediate solution, it would substantially help clarifying the comparison between theory and experiment.

Furthermore, a study of $B_s^0\to D_{s 2}^{\ast -}\pi^+$  would be an important test of the consistency between theory and experiment as far as $\tau_{3/2}(1)$ is concerned.

\subsection{The two narrow $j=1/2$ $D_{sJ}$ states}
\label{sec:Ds}

It is very fortunate that the strange $j=1/2$ $D^{\ast\ast}$-states, $D_{s 0}^\ast (2317)$ and $D_{s 1}(2460)$, are very narrow, because their masses are below their respective  $D^{(*)} K$ thresholds. The broad non-strange states are heavier than the $D^{(*)} \pi$ threshold.  While the $SU(3)$ symmetry breaking is large in the phase space, it can still be expected to work well for the electroweak amplitudes and strong couplings, as has been observed most often. 

The narrowness of the states offers an exceptional possibility to test the theoretical predictions in a much better experimental situation. It eliminates at the same time the problem of the non-resonant background, and  interference with competing crossed channels, since both should be relatively negligible near the peak.

The effect of $SU(3)$ breaking is expected to be small for the lowest lying states with given quantum numbers.
 We therefore  expect $\tau_{1/2}(1)$ to be rather close to the non-strange case. Note that in the lattice QCD study of ref.~\cite{wagner09}, no significant dependence of $\tau_{1/2}(1)$ on the light quark mass has been observed.~\footnote{
 A proposal to study the $B_s\to D_{sJ}$ transition has been made in ref.~\cite{huang} in order to test whether or not the $D_{sJ}$ states are indeed the ``$\bar q q$" structures. They use the QCD sum rule calculations in HQET and find huge SU(3) breaking effect ($\sim 100\%$) in the form factor [compare eq.~(34) in ref.~\cite{dai} with eq.(32) in ref.~\cite{huang}], which contradicts the lattice QCD findings of ref.~\cite{wagner09}.} 

We should emphasize once again a great advantage of the non-leptonic over semileptonic $B_s^0$-decays in that they do not have the neutrino identification problem, but have the two body final state with well known masses.
Theoretically,  $B_s^0 \to \overline{D}_{sJ}~\pi$ is the most interesting decay because it is described by the pion emission diagram only ($B_s$ annihilation being neglected as usual). In the factorization approximation, it directly yields $\tau_{1/2}(1)$.

{\sl Warning concerning a possible misinterpretation of $D_s(2317,2460)$}:  A potential caveat concerning the  $D_{s 0}^\ast (2317)$ and $D_{s 1}(2460)$ is that they might not be the ``$q \bar q$" states. A controversy resides in the fact that the measured masses of these states are lower than predicted. However, the level ordering of the ``$q \bar q$" states,  $0^-, 1^-, 0^+, 1^+, 1^+, 2^+$, is consistent with what is observed with the $D_{sJ}$ mesons so far. Moreover, the study of their transition properties does not favor an exotic assignment either. We must underline that a measurement of the decays proposed here will also provide an extra check of the $q\overline{q}$ structure of  $D_s(2317,2460)$.

\section{Decay branching fractions of $D_{s0}^{*}(2317)^+$ and 
$D_{s1}(2460)^+$ states}

Of course, to measure the $B_s^{0}\to \overline{D}_{sJ}^+ \pi^-$ rates, the knowledge of the $D_{sJ}$ branching ratios is necessary. 

In ref.~\cite{ref:pdg} only absolute values for the $D_{s1}^+(2460)$  
branching fractions are quoted. This is because, at present,
there is only a single measurement \cite{ref:babara} of the $D_{s1}^+(2460)$
production in $B \rightarrow D_{s1}^+(2460) \overline{D}^{(*)}$ decays,
independently of the decay channel for the $D_{s1}^+(2460)$.
Production of $D_{sJ}$ states was studied by considering
the missing mass distribution in  $B \rightarrow \overline{D}^{(*)} X$ decays
and
signals were observed only for $X=D_s^+,~D_s^{*+}~{\rm and}~D_{s1}^+(2460)$.
As a result, there is no absolute decay branching fraction measurement
for the $D_{s0}^*(2317)$.

\subsection{$D_{s0}^{*}(2317)^+$ decay channels}
Experimental results collected in \cite{ref:pdg} are reminded in 
Table \ref{tab:ds0star}.

\begin{table}[!htb]
\begin{center}
  \begin{tabular}{|c|c|c|}
    \hline
 Decay channel  & $90\%$ C.L. limit & comment\\
\hline
 $D_s^+ \gamma$ & $<0.05$ & forbidden\\
 $D_s^{*+} \gamma$ & $<0.059$& allowed\\
 $D_s^+ \gamma \gamma$ & $<0.18$ & allowed\\
 $D_s^{*+} \pi^0$ & $<0.11$& forbidden\\
 $D_s^+ \pi^+\pi^-$ & $<0.004$& forbidden\\
 $D_s^+ \pi^0 \pi^0$ & $<0.25$& forbidden\\
\hline
  \end{tabular}
  \caption[]{\small {$90\%$ C.L. limits on branching fractions
for different decay channels measured relatively to the $D_s^{+} \pi^0$
channel. In the last column are indicated the allowed and forbiddden decay
channels from angular momentum and parity conservation. } 
  \label{tab:ds0star}}
\end{center}
\end{table}

The electromagnetic $D_s^+ \gamma \gamma$ is expected to be negligible as
two photons have to be radiated. It thus remains only two possible decay
channels for the $D_{s0}^{*}(2317)^+$. In Table \ref{tab:ds0star_th}
are indicated some model expectations on these decay channels
\cite{ref:bardeen,ref:godfrey}.

\begin{table}[!htb]
\begin{center}
  \begin{tabular}{|c|c|c|c|}
    \hline
 Decay channel  & model 1 \cite{ref:bardeen}   & model 2 \cite{ref:godfrey}& $90\%$ C.L. limit \\
\hline
 $D_s^+ \pi^0$ & $92.5\%$ & $84\%$ &\\
 $D_s^{*+} \gamma$ & $7.5\%$ & $16\%$ & $<0.059$\\
\hline
  \end{tabular}
  \caption[]{\small {Some model expectations for $D_{s0}^{*}(2317)^+$ 
branching fractions compared with the experimental result.} 
  \label{tab:ds0star_th}}
\end{center}
\end{table}
The present limit on the $D_s^{*+} \gamma$ decay channel is more stringent than
the estimates; in the following we will use:

\begin{equation} 
 {\cal B}(D_{s0}^{*}(2317)^+ \rightarrow D_s^+ \pi^0)= (97 \pm 3)\%. 
\end{equation}

\subsection{$D_{s1}(2460)^+$ decay channels} 

Experimental results collected in \cite{ref:pdg} are reminded in 
Table \ref{tab:ds1}.

\begin{table}[!htb]
\begin{center}
  \begin{tabular}{|c|c|c|}
    \hline
 Decay channel  & value or limit & comment\\
\hline
 $D_s^{*+} \pi^0$ & $(48\pm11)\%$ & allowed\\
 $D_s^+ \gamma$ & $(18\pm4)\%$& allowed\\
 $D_s^+ \pi^+\pi^-$ & $(4.3\pm1.3)\%$& allowed\\
 $D_s^{*+} \gamma$ & $<0.08$ (90$\%$ C.L.) & allowed\\
 $D_{s0}^{*}(2317)^+ \gamma$ & $(3.7^{+5.0}_{-2.4})\%$& allowed\\
 $D_s^+ \pi^0$ & $<0.042$ (95$\%$ C.L.)& forbidden\\
 $D_s^+ \pi^0 \pi^0$ & $<0.68$ (95$\%$ C.L.)& allowed\\
 $D_s^+ \gamma \gamma$ & $<0.33$ (95$\%$ C.L.)& allowed\\
\hline
  \end{tabular}
  \caption[]{\small {Measured branching fractions
or upper limits 
for different $D_{s1}(2460)^+$ decay channels. 
In the last column are indicated the allowed and forbiddden decay
channels from angular momentum and parity conservation. } 
  \label{tab:ds1}}
\end{center}
\end{table}

Many decay channels are possible and individual decay branching fractions
are not accurately measured. The situation is thus experimentally less
favourable than for the $D_{s0}^{*}(2317)^+$ resonance to measure the
production rate of this state.

\section{Expected rates at LHCb}

\subsection{Analysis method}
We would like to have a measurement of the decay chain 
$B_s^0 \rightarrow D_{s0}^{*-} \pi^+,~D_{s0}^{*-} \rightarrow D_s^- \pi^0,~
D_s^-  \rightarrow K^+ K^- \pi^-$  in which the $\pi^0$ meson
cascading from the $D_{s0}^{*-}$ is not detected.

It is proposed to measure the missing $\pi^0$ 4-momentum using the
measurement of the $B_s^0$ direction and two mass constraints ($m_{\pi^0}$
and $m_{B_s^0}$). The $B_s^0$ direction is determined from the reconstructed
positions of the $pp$ interaction and the $B_s^0$ decay vertices.
Measured uncertainties on these quantities can be included in a fit
with the two mass constraints.

There could be 2 solutions for the signal and a study based on simulated
events may help to choose one of these possibilities. The amount of background
candidates can be decreased using the fit $\chi^2$ probability.

For signal events, as the $D_{s0}^{*-}$ has a very small
intrinsic width, one expects to observe a  peak in the 
$D_s^- \pi^0$ mass distribution having a width which depends mainly
on the accuracy of tracking capabilities.

\subsection{Expected rates}\label{sec:rates}

The proposed analysis is based on the same charged particles final state
which was already measured in LHCb for the channel:
$B_s^0 \rightarrow D_{s}^{-} \pi^+,~D_s^-  \rightarrow K^+ K^- \pi^-$.
Few selection criteria have to be removed to allow for the missing
$\pi^0$ meson and in particular the condition on the similarity
between the directions defined by the two vertices and by the 
$K^+ K^- \pi^+ \pi^-$ momentum.

Analyzing 336 $pb^{-1}$ integrated luminosity, LHCb has measured
\cite{ref:lhcbbstodspi}
about 6000 $B_s^0 \rightarrow D_{s}^{-} \pi^+$ decays. The number
of $B_s^0 \rightarrow D_{s0}^{*-} \pi^+$ reconstructed events
can be estimated by comparing the corresponding branching fractions
for the two decay channels.

From $SU(3)$ symmetry and factorization, we can simply identify the branching fraction of $B_s \to \overline{D}_{sJ}\pi$ with the one of the neutral $B$ into charged $\overline{D}^{\ast\ast}$ and $\pi$. Indeed, the phase space is also very close to the one in the non strange case. In view of the other uncertainties, we can safely disregard any $SU(3)$ effect.
This means from the measured case, the one of $J^P=0^+$,
\begin{equation}
{\cal B}(B_s^0 \rightarrow D_{s0}^{*-}(2317) \pi^+) = (1.0  \pm 0.5) \times 10^{-4}\,,
\end{equation}
where we average the results of Belle and BaBar for the non-strange decays (BaBar is presently not published). This value agrees with  the theoretical expectation using the heavy quark limit [$10^{-4}$, c.f. eq.~(\ref{NL0plus})]. 
However, using the experimental value for the non-strange decays together with the $SU(3)$ light flavor symmetry is likely to be better than the result derived in the heavy quark limit and assuming exact factorization.

To assess the soundness of the $SU(3)$ assumption, let us consider the decays to $D,D_s$. The LHCb collaboration has measured:
\begin{equation}
{\cal B}(B_s^0 \rightarrow D_{s}^{-} \pi^+) = (2.95 \pm 0.28) \times 10^{-3}.
\end{equation}
In this expression we have added in quadrature the different uncertainties
quoted in the publication. The value agrees well, as expected, with the 
corresponding measurement for the ${B}_d^0$ meson.
\begin{equation}
{\cal B}(B_d^0 \rightarrow D^{-} \pi^+) = (2.68 \pm 0.13) \times 10^{-3}.
\end{equation}

Analyzing an integrated luminosity of 1$fb^{-1}$, the LHCb collaboration
can thus expect to reconstruct:
\begin{equation}
{\cal N}(B_s^0 \rightarrow D_{s0}^{*-}(2317) \pi^+) = 1800 \times {1\over 3}\times (1 \pm 1/2)
\times {\cal B}(D_{s0}^{*-} \rightarrow D_s^- \pi^0) 
\times \epsilon_{\pi^0},
\end{equation} 
with the $D_s^-$ meson reconstructed in the $K^+ K^- \pi^-$ decay channel.
The quantity  $\epsilon_{\pi^0}$ corresponds to the efficiency of the 
additional cuts which have to be applied to select the events.

A very few hundred of events are expected and
the signal visibility will thus depend mainly on the mass resolution
for the $D_s^- \pi^0$ system and on the combinatorial background level.

\vskip1cm
\begin{huge} \bf                      \end{huge}

\section{Conclusion}
\subsection{Feasibility of the proposal}
\label{sec:feasibility}

We propose an experimental study of the $B_s\to \overline D_{sJ}\pi$ decays that would provide us with an important verification of the observations made in the corresponding non-strange modes. 
Furthermore it would allow us to elucidate the problem of small value of $\tau_{1/2}(1)$.

If a really unexpected value for ${\cal B}(B_s^0 \to D_{s0}^{*-} \pi^+)$ is found,  this could mean that 
\begin{itemize}
\item[--] either we are mistaken in the theoretical evaluation of $\tau_{1/2}(1)$, which would be very surprising in view of good consistency of several approaches, or the $1/m_c$ corrections are exceedingly large in the $j=1/2$ case, 
\item[--] or the narrow $D_{sJ}$-states situated below the $D^{(\ast)} K$ thresholds are not the ``$q \bar q$" states  with $j=1/2$ (ree ref.~\cite{DsJ} for a review).
\end{itemize}
Both these possibilities do not seem plausible to us. 
The remaining uncertainty on the theoretical side could be significantly reduced by the lattice study of the $B_s^0 \to \overline{D}_s^{\ast\ast}$ transition form factors at finite heavy quark masses.

If the expected rate is confirmed, that would set beyond doubt the theoretical estimates of small values for $\tau_{1/2}(1)$ and it would confirm the assignment of the $D_{sJ}$ states. 
A strong suspicion would be confirmed against the semileptonic measurements or identifications of resonances performed in the non-strange case.

\subsection{Remaining problems on the non-strange side}
Even if the answer of the proposed experiment is in agreement with  theoretical expectations made by adopting the ``$q \bar q$" assignment to the $D_s^{\ast\ast}$ states, it will still not give us the full explanation to the problems observed in the non-strange case. The problems encountered on the experimental side, especially in semileptonic non-strange decays, remain to be understood: the origin of the discrepancy between Belle and BaBar; why so large apparent rates for decay to $0^+$? A theoretical explanation for the large number of events in the non-strange semileptonic decay is missing.

The observed excess of events in $D^{(\ast)} \pi$ (around $1 \%$) and in $Dn \pi$, that in our opinion are not  the lowest $j=1/2$ or $j=3/2$ states, needs an explanation. Such events should have their counterpart in non-leptonic decays. To test an excess in the $D \pi$ channel, a  study of the decay $B_d^0 \to \overline{D}^0 \pi^- \pi^+$ at LHCb would be very welcome.~\footnote{
To interpret these events, one could think of a possible contribution from radial excitations in the non-strange sector, considered already in ref.~\cite{memorino2}, and strongly advocated in ref.~\cite{ligeti}. However, one must note the following. Calculating the semileptonic transition rate from $B$ to the first radial excitation of the $\overline{D}^{(\ast)}$ within the same approach as for the orbital excitations above~\cite{morenas}, in the heavy quark limit, we find a very small number with respect to the decay to the ground states, $\simeq 0.01$. This is because the corresponding Isgur-Wise function is very small, reaching its maximum at $w_{\rm max} \simeq 1.3$, with $\xi(w_{\rm max}) \simeq 0.1$. Such a small number agrees with the findings made by using lattice QCD at $w$ close to $1$~\cite{hein}.  The contribution to the non-leptonic decay should then also be small. This seems to discard the radial excitation interpretation of the remaining events.}

\newpage 
\section*{Acknowledgements} We would like to thank R.~Aleksan, B.~Blossier and O.~P\`ene for useful information and discussions.

\end{document}